\documentclass[a4paper,aps,pra,
,floatfix,twocolumn,nofootinbib]{revtex4-1}

\usepackage{bm}			    
\usepackage{amsmath} 
\usepackage{amssymb}
\usepackage{latexsym}
\usepackage{amsfonts}
\usepackage{color}
\usepackage{graphicx}

\newcommand{\ket}[1]{\vert{#1}\rangle} 
\newcommand{\bra}[1]{\langle{#1}\vert} 
 
\newcommand{\mean}[1]{\langle #1 \rangle}

\DeclareMathOperator{\Tr}{Tr}

\newcommand{\abs}[1]{\left|#1\right|} 

\renewcommand{\vec}[1]{\mathbf{#1}}

\renewcommand\Im{\operatorname{Im }}

\newcommand{\beq}{\begin{equation}}
\newcommand{\eeq}{\end{equation}}

\newcommand{\bc}{\begin{center}}
\newcommand{\ec}{\end{center}}

\newcommand{\be}{\begin{equation}}
\newcommand{\ee}{\end{equation}}

\newcommand{\pare}[1]{\left( #1 \right)}
\newcommand{\key}[1]{\left\{ #1 \right\}}
\newcommand{\ben}{\begin{eqnarray}}
\newcommand{\een}{\end{eqnarray}}

\begin{document}

\title{Quantum transport efficiency and Fourier's law}


\author{Daniel Manzano$^{1,2,3}$}
\email{daniel.manzano@uibk.ac.at}
\author{Markus Tiersch$^{1,2}$}
\author{Ali Asadian$^{1}$}
\author{Hans J. Briegel$^{1,2}$}

\affiliation{$^{1}$Institute for Theoretical Physics,
University of Innsbruck,
Technikerstr.~25, A-6020 Innsbruck, Austria, Europe}
\affiliation{$^{2}$Institute for Quantum Optics and Quantum Information,
Austrian Academy of Sciences,
Technikerstr.~21A, A-6020 Innsbruck, Austria, Europe}
\affiliation{$^{3}$Instituto Carlos I de Fisica Teorica y Computacional,
University of Granada,
Av.~Fuentenueva s/n, 18071 Granada, Spain, Europe}

\begin{abstract}
We analyze the steady-state energy transfer in a chain of coupled two-level systems connecting two thermal reservoirs. Through an analytic treatment we find that the energy current is independent of the system size, hence violating Fourier's law of heat conduction.
The classical diffusive behavior in Fourier's law of heat conduction can be recovered by introducing decoherence to the quantum systems constituting the chain.
We relate these results to recent discussions of energy transport in biological light-harvesting systems, and discuss the role of quantum coherence and entanglement.
\end{abstract}

\maketitle

\section{introduction}

In recent years, energy propagation in systems that must be described in a quantum 
mechanical way has become a growing field. This growth is partially due to the fact that the understanding of how energy flow can be controlled 
and efficiently distributed has been identified as one of the crucial fields of study for the development of modern societies \cite{usdoe:09,dubi:rmp11}.
One of the conceptual pillars in energy transport, the validity of Fourier's law of heat conduction, has become an active area of investigation and has 
been investigated in classical~\cite{garrido:prl01,dhar:ap08} and quantum systems~\cite{saito:epl03,michel:prl05,dubi:pre09}. 

Since experimental evidence for quantum coherent excitation transport in the early light-harvesting step of photosynthesis has been presented~\cite{engel:nature07,lee:science07}, investigations in systems of molecular biology have focused on the question, to what extent quantum mechanics contributes to the near perfect transport efficiency in light-harvesting. The emphasis has been put on the \emph{transient} transport efficiency of an initial excitation in the presence of noise and disorder~\cite{aspuru,plenio:njp10,scholak:pre11}. The experiments have been performed with pulsed femtosecond laser sources to excite and probe the molecule samples, whereas it has been suggested~\cite{Brumer,Mancal:NJP} that the light-harvesting process \emph{in vivo} would be described more accurately in a steady-state scenario, because the light flux coming from the sun is essentially static on time scales that are relevant for molecular excitation transport. A realistic treatment of the energy transport through photosynthetic complexes in such a scenario will be a formidable task and is yet to be developed. In this paper, we will reconsider the treatment of the light-harvesting complex as a system of coupled two-level systems \cite{aspuru,plenio:njp10,scholak:pre11,Mancal:NJP} and study the role of noise and entanglement in a steady-state scenario. We will concentrate here on a simple one-dimensional model, for which we find an analytical formula for the heat current and its dependence on the dephasing. Using this formula, we find a transition from ballistic to diffusive transport due to decoherence, recovering earlier results \cite{Znidaric_11}. We also discuss implications regarding the possible role of the environment for the transport efficiency \cite{aspuru,plenio:njp10,scholak:pre11} and the occurrence of entanglement in the steady state.

An important step in the understanding how Fourier's law emerges from the quantum domain has been made by Michel {\it et al}.~\cite{michel:prl05}.
In this work Fourier's law is derived for a model system that is a chain of $N$ identical coupled subunits, where each of the subunits has a single ground state and a narrow ``band'' of equally spaced excited states.
In the present work, we employ a similar system, i.e., a one-dimensional chain of two-level systems, for which we compare the energy current in 
the classical analog, where Fourier's law applies, with the quantum case, where we find the energy current to be independent of the 
chain length. This means that for the one-dimensional chain of two-level atoms Fourier's law applies for the classical variant but there is a distinct violation in the quantum transport scenario.
By introducing dephasing to the quantum model, we can study the transition from coherent to incoherent transport and show how Fourier's law can be recovered from the quantum case.

Fourier's law of heat conduction states that the heat current through a classical macroscopic object is proportional to the applied temperature gradient~\cite{fourier_22},
\beq
\vec J  = -\kappa \boldsymbol\nabla T,
\eeq
where $\kappa$ is the thermal conductivity. For a one-dimensional homogeneous object, the heat current is therefore determined by the temperature difference of the two heat baths $\Delta T$, and the object length $L$.
Generally, the validity of Fourier's law does not seem to be strictly linked to the classical or quantum nature of the system.
For example, in the classical limit, for diffusive systems Fourier's law can be applied, but for ballistic systems
in one and two dimensions there are divergences of the thermal conductivity as $\kappa\sim L^\alpha$
 (see~\cite{dhar:ap08} for a review of heat transfer in low dimensional systems).
For a discretized object composed of $N$ equally spaced parts (sites), $L\propto N$ and thus
\beq \label{eq:fourier}
J =- \kappa \frac{\Delta T}{L} = -c N^\alpha \frac{\Delta T}{N} = -c \, \Delta T \, N^{\alpha-1},
\eeq
where $c$ is a constant of proportionality.
For some one-dimensional quantum systems, on the other hand, there is evidence that Fourier's law is valid, i.e., $\alpha=0$~\cite{saito:epl03,michel:prl05}.

\section{quantum model}

\begin{figure}
	\centering
	\includegraphics[width=0.7\linewidth]{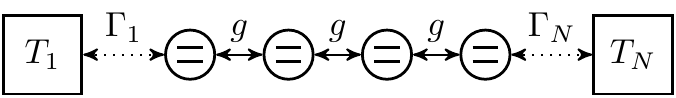}
	\caption{Chain of two-level quantum systems with its terminal sites coupled to heat baths of different temperatures.}
	\label{fig:system}
\end{figure}
The quantum system considered is a one-dimensional chain of $N\ge2$ two-level systems with coherent next-neighbor couplings as depicted in Fig.~\ref{fig:system}.
The Hamiltonian is
\beq \label{eq:hamiltonian}
H= \sum_{k=1}^N \frac{\hbar\omega}{2} \sigma_k^z + \sum_{k=1}^{N-1} \hbar g \left(\sigma^+_k \sigma^-_{k+1} + \sigma^-_k \sigma^+_{k+1} \right),
\eeq
where $\sigma^z_k$, $\sigma^+_k$, and $\sigma^-_k$ are the Pauli-$z$, raising, and lowering operators in the basis of ground and excited states of the $k$th two-level system, respectively, with on-site energy $\hbar\omega$ and coupling strength~$g$.
Similar simple models of coupled effective two-level systems are used in recent analyses of energy transfer in photosynthetic complexes~\cite{plenio:njp10,aspuru,scholak:pre11}
and spin transport in models for magnetism~\cite{prosen:prl08,Znidaric_11,krevski:prl09}. 
The influence of the two heat baths is modeled by incoherently coupling each of the terminal sites to a bosonic heat bath described by a master equation of Lindblad form.
The system dynamics is then described by the master equation
\beq \label{eq:ME}
\dot{\rho} = -\frac{i}{\hbar} [H,\rho] + \mathcal{L}_1\rho + \mathcal{L}_N\rho,
\eeq
where $\mathcal{L}_k$ acts on the first (last) site for $k=1\,(N)$, respectively, and is given by
\begin{align}\label{eq:lindblad}
\mathcal{L}_k \rho
&=
\Gamma_k (n_k+1) \left( \sigma^-_k \rho \sigma^+_k -\frac{1}{2} \left\{\sigma^+_k\sigma^-_k,\rho\right\} \right) \nonumber \\
&+
\Gamma_k n_k \left( \sigma^+_k \rho \sigma^-_k -\frac{1}{2} \left\{\sigma^-_k\sigma^+_k,\rho\right\} \right).
\end{align}
The first term in $\mathcal{L}_k$ accounts for emission into the reservoir, the second term accounts for absorption, $\Gamma_k$ is the interaction rate, and $n_k=1/\left\{ \exp[\hbar\omega/(k_B T_k)] - 1\right\}$ is the temperature-dependent mean excitation number at the resonance frequency in the respective bosonic thermal reservoir~\cite{Breuer}, with $k_B$ being Boltzmann's constant.

\section{heat current}

The expression of the heat current for a quantum system, $J_Q$, is derived from the time-derivative of the energy of the system,
\beq
\dot{E}= \frac{d}{dt} \mean{H} = \Tr \left( H \dot\rho \right) = 0,
\eeq
which vanishes in the steady state.
When inserting~\eqref{eq:ME} into this expression, we obtain
\beq
0 = \Tr \left( H \mathcal{L}_1 \rho + H \mathcal{L}_N \rho \right) =: J_1 + J_N,
\eeq
on the basis of which one can define the heat current to and from the respective reservoirs, both being of opposite sign, but equal in magnitude~\cite{Breuer}. 
The heat current through the chain is therefore equal to the net energy that enters the network from one reservoir and exits to the other per unit time, 
i.e., the quantity $J_Q=\abs{J_1}=\abs{J_N}$.
A straightforward evaluation of $J_Q$ for our system in the steady state yields the compact expression
\beq \label{eq:heatcurrent}
J_Q = \gamma_1 \hbar \omega \big(s_1 - \mean{\sigma^+_1\sigma^-_1}\big) 
- \frac{\gamma_1 \hbar g}{2} \big(\mean{\sigma^+_1 \sigma^-_2} + \mean{\sigma^-_1 \sigma^+_2}\big),
\eeq
where $\gamma_1=\Gamma_1(2n_1+1)$ denotes the effective coupling to the reservoir, $s_1=n_1/(2n_1+1)$ is the excited-state population of a single two-level system in thermal equilibrium with reservoir~1, and all expectation values are taken with respect to the steady state of the chain.
The heat current in the steady state is thus solely characterized by the excited-state population of the first site and its specific energy gap, and since $\mean{\sigma^+_1 \sigma^-_2} = \mean{\sigma^-_1 \sigma^+_2}^*$, it is furthermore given by the \emph{real} part of the coherence between sites $1$ and $2$.
An analogous expression can be given for the last site of the chain, which is connected to the second heat bath.


For the complete expression of the heat current, we need the excited-state population of the first site, $\mean{\sigma^+_1\sigma^-_1}$, and the coherences between the first two sites, $\mean{\sigma^+_1 \sigma^-_2}$.
The excited-state populations of the individual sites in the steady state can be obtained from considering specific matrix elements of the master equation of the kind $\frac{\partial}{\partial t}\mean{\sigma^+_k \sigma^-_k} = \Tr(\sigma^+_k \sigma^-_k \dot{\rho})=0$.
There are different cases: sites~1 and $N$, which are connected to their respective heat baths, and the remaining sites, which are in the middle of the chain.
The relevant equations for the terminal sites $k=1$ and $k=N$ yield:
\begin{align*} \label{eq:MEpop1}
\gamma_1 \big( s_1 - \mean{\sigma^+_1 \sigma^-_1} \big) &= ig \big( \mean{\sigma^+_1 \sigma^-_2} - \mean{\sigma^-_1 \sigma^+_2}\big), \\
\gamma_N \big( s_N - \mean{\sigma^+_N \sigma^-_N} \big) &= -ig \big( \mean{\sigma^+_{N-1} \sigma^-_N} - \mean{\sigma^-_{N-1} \sigma^+_N}\big). \nonumber
\end{align*}
For the inner sites, $1<k<N$, we obtain
\begin{equation} \label{eq:MEcoh}
\mean{\sigma^+_{k-1} \sigma^-_k} - \mean{\sigma^-_{k-1} \sigma^+_k} = \mean{\sigma^+_k \sigma^-_{k+1}} - \mean{\sigma^-_k \sigma^+_{k+1}},
\end{equation}
that is, the \emph{imaginary} parts of all coherences between neighboring sites are equal.
These equations motivate the following general form for the excited-state populations of the terminal sites:
\begin{align}	\label{eq:pops}
\mean{\sigma^+_1 \sigma^-_1} &= s_1 -\Delta/\gamma_1, &
\mean{\sigma^+_N \sigma^-_N} &= s_N + \Delta/\gamma_N.
\end{align}
The transport along the chain thus causes a shift of the excited-state population of the terminal sites from the thermal equilibrium by $\Delta/\gamma_k$, where $\Delta = ig \big( \mean{\sigma^+_1 \sigma^-_2} - \mean{\sigma^-_1 \sigma^+_2}\big)$.
The coherences, and thereby $\Delta$, can be obtained by a similar argument.
Summing up coherences of the steady state,
$\frac{\partial}{\partial t}\sum_{k=1}^{N-1} \mean{\sigma_k^+\sigma_{k+1}^-} = 0$,
provides the equation
\begin{equation*}
-ig \left( \mean{\sigma_1^+\sigma_1^-}-\mean{\sigma_N^+\sigma_N^-} \right) = \frac{\gamma_1}{2}\mean{\sigma_1^+\sigma_2^-} + \frac{\gamma_N}{2}\mean{\sigma_{N-1}^+\sigma_N^-},
\end{equation*}
the imaginary part of which, using \eqref{eq:MEcoh} and \eqref{eq:pops}, yields
\beq
\Delta = \frac{4 g^2 \gamma_1\gamma_N (s_1-s_N)}{(\gamma_1+\gamma_N)(4g^2+\gamma_1\gamma_N)}.
\eeq

Next, for the complete solution of the heat current, we need the real part of the next-neighbor coherences, i.e., $\mean{\sigma_1^+\sigma_2^-}+\mean{\sigma_1^-\sigma_2^+}$. For a chain of sites with uniform on-site energy as considered here, these coherences are purely imaginary, which can be shown from the structure of the master equation (see Appendix A).
We thus arrive at the final expression of the heat current for a uniform quantum chain,
\beq \label{eq:finalJ}
J_Q = \hbar \omega \Delta = -2 \hbar \omega g \Im \mean{\sigma_1^+\sigma_2^-}.
\eeq


Here, we observe an important property of the heat current. It is independent of the chain length and thus violates Fourier's law, that is, the thermal conductivity scales as $\kappa\sim N$, and thus $\alpha=1$.


\begin{figure}
	\centering
	\includegraphics[width=0.45\linewidth]{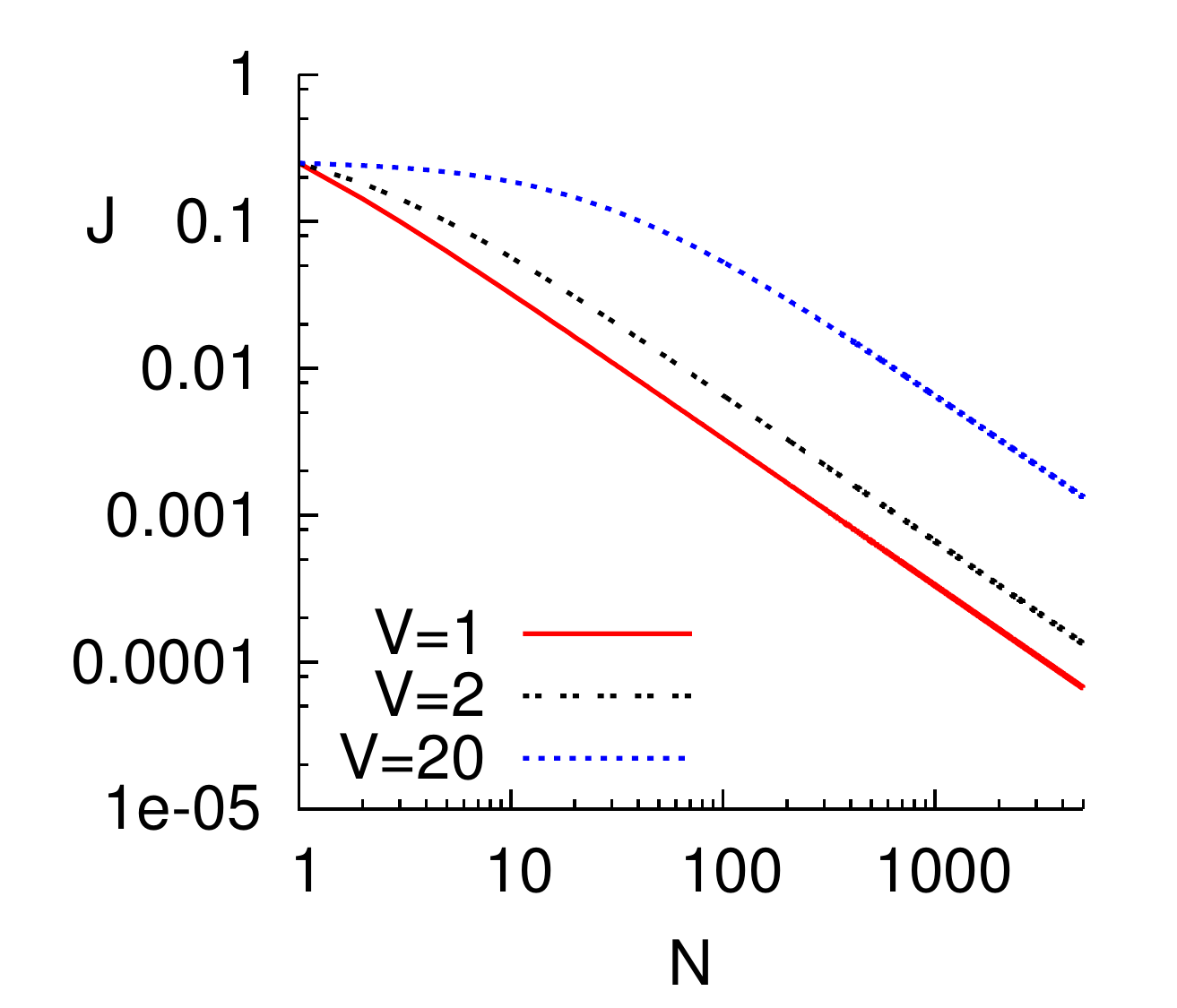}\hfill%
	\includegraphics[width=0.45\linewidth]{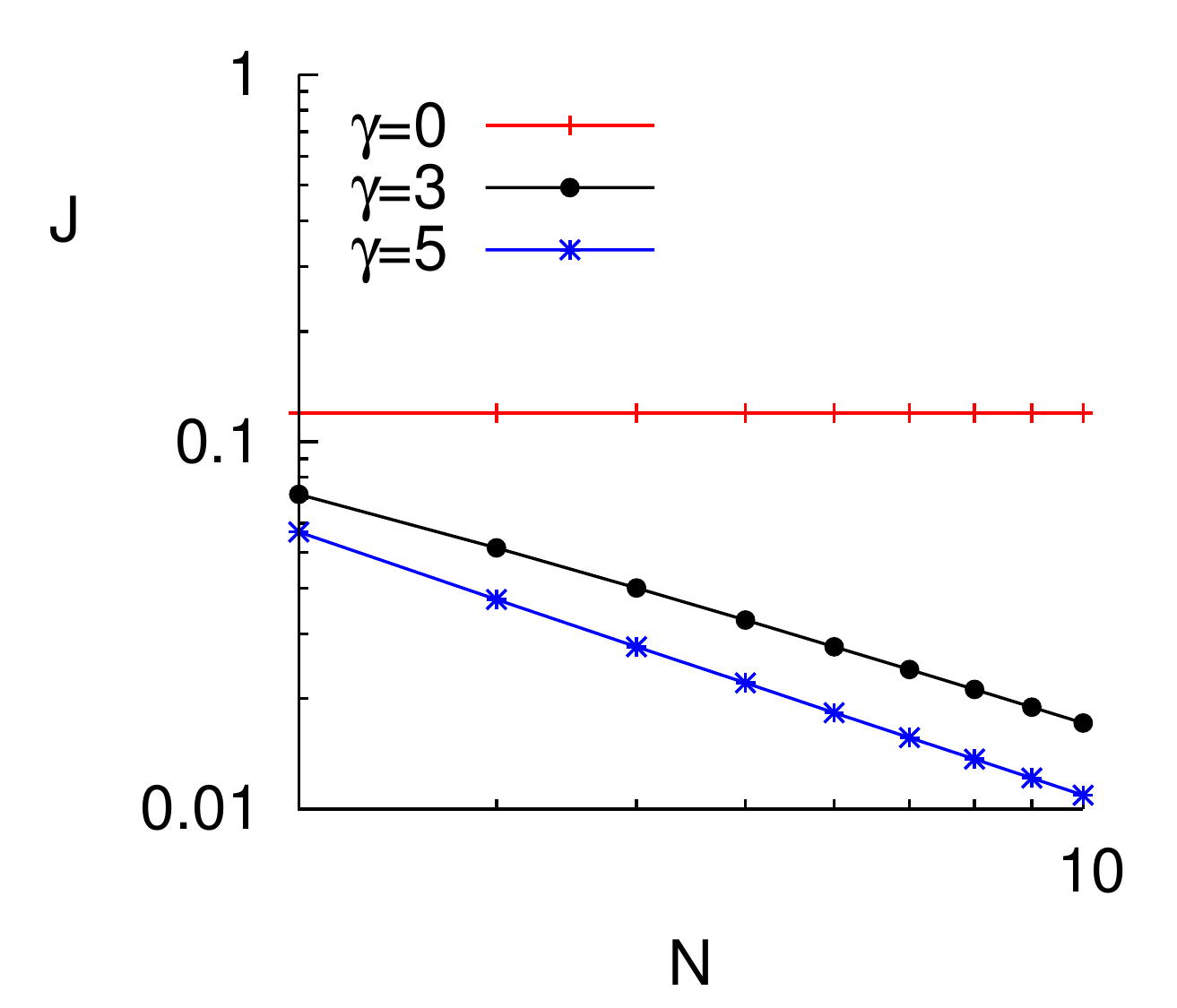}
	\caption{(Color online) Heat current as a function of the system size in a log-log plot (left) for the classical model for different values of the coupling~$V$, and (right) for the quantum model for different values of the dephasing rate~$\gamma$. Unless otherwise indicated $V=1$, $g=1$, $k_B T_1=1$, $\hbar\omega=1$, $\Gamma_1=\Gamma_N=1$, and $T_N=0$.}
	\label{fig2}
\end{figure}

\section{Classical model}

Next, for comparison of the quantum model with the analogous classical model, we derive the heat current for the latter, which corresponds to the symmetric simple exclusion process~\cite{derrida:jsp02}, or F\"{o}rster-type hopping~\cite{MayKuehn}. It is a chain of $N$ sites, which each may carry a single particle (excitation) that probabilistically moves between neighboring sites. The probability of
the particles to jump to each neighboring site are equal, with the only condition that each site can carry only one particle.
This diffusive model fulfills Fourier's law. 
The classical probability for a particle to be at site $k$ is given by $P_k$, and the state of the system at any given time is defined 
by the set of probabilities $\left\{ P_k \right\} $ for each site. As in the quantum model, the extreme sites are connected with thermal baths.
The master equation~\eqref{eq:ME} is thus turned into a Pauli master equation, i.e., a set of classical rate equations:
\begin{align*}
\dot{P}_1	=& \Gamma_1 n_1+P_1\big(-\Gamma_1(n_1+1)-\Gamma_1 n_1-V\big) + V P_2 \nonumber\\
\dot{P}_{k}	=& V\pare{P_{k-1}+P_{k+1}-2P_k} \qquad (k\neq1,N) \\
\dot{P}_N	=& \Gamma_N n_N+P_N\big(-\Gamma_N(n_N+1)-\Gamma_N n_N-V\big) + V P_{N-1} \nonumber
\end{align*}
where $V$ is the constant rate of hopping between sites, and $\Gamma _k$  and $n_k$ are the bath parameters, with the same interpretation as in the quantum master equation.
In the classical system, the heat current is defined by %
$
J_C=\left| V\pare{P_{i+1}-P_{i}}\right|,
$
i.e., the net transfer rate of energy between sites,
which in the steady state yields
\begin{equation*}
J_{C}=\frac{\gamma_1\gamma_N V (s_1-s_N)}{V (\gamma_1 + \gamma_N) + \gamma_1 \gamma_N (N-1)}.
\end{equation*}
In the limit $N\to \infty$, the heat current scales with the system size as
$
J_C \sim V(s_1-s_N)/N.
$
Therefore, the heat current of the classical analog obeys Fourier's law with $\kappa =\text{const.}$ and $\alpha=0$, in contrast to the quantum system.


\begin{figure}
	\includegraphics[width=0.45\linewidth]{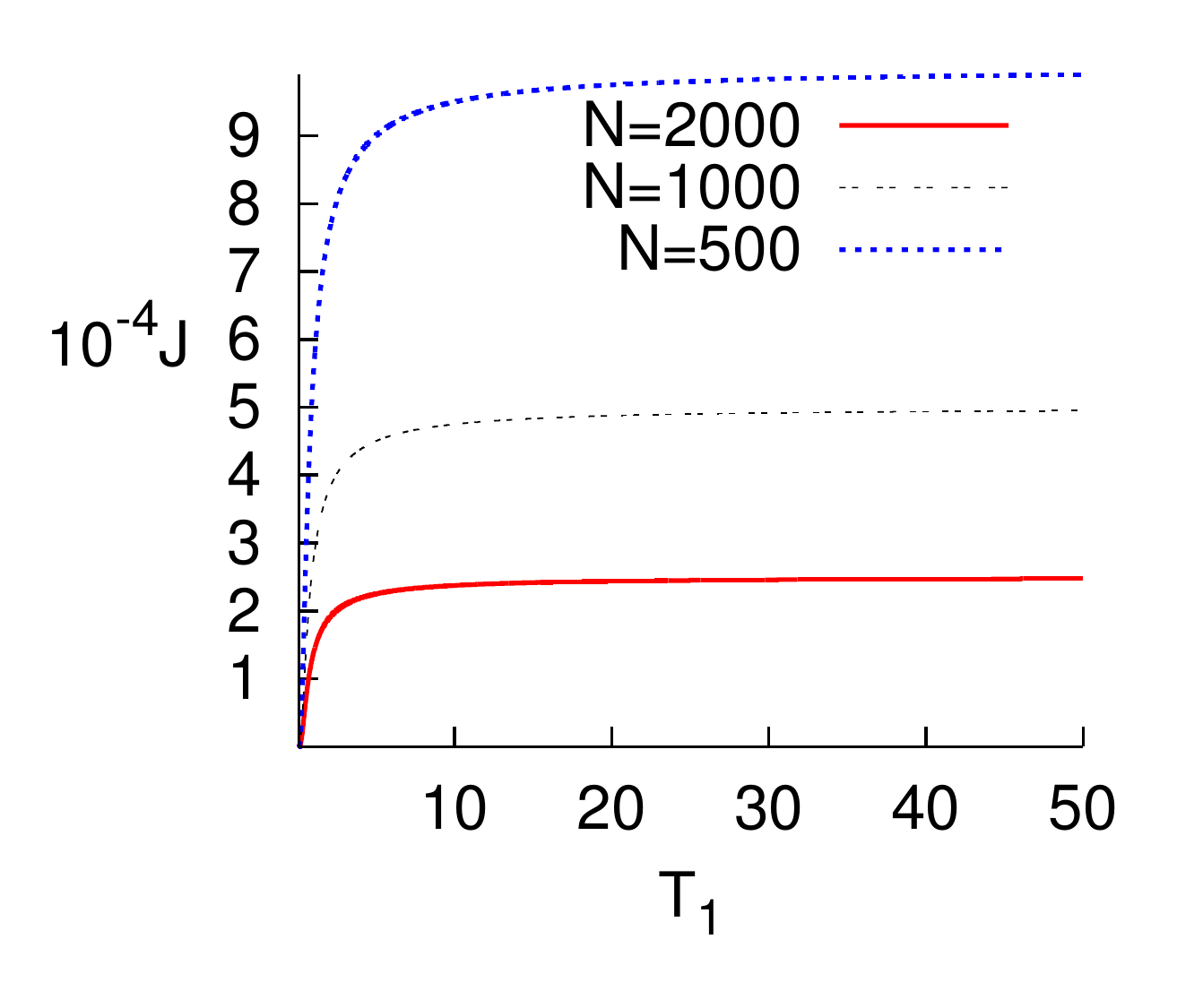}\hfill%
	\includegraphics[width=0.45\linewidth]{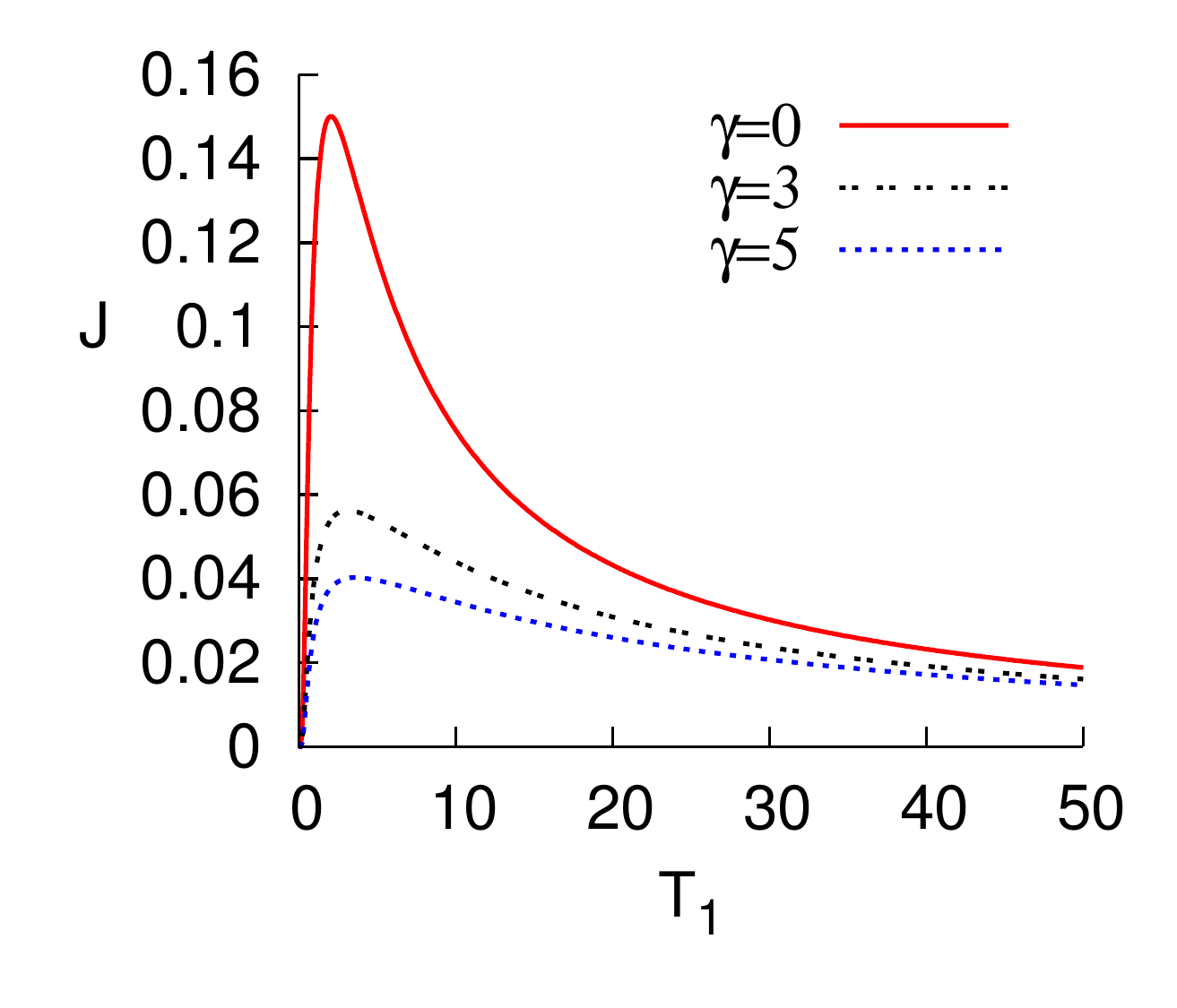}
	\caption{(Color online) Heat current as a function of temperature $T_1$ with $T_N=0$ fixed, for the classical transport for different sizes of the system (left), and the quantum transport for $N=4$ and different dephasing rate (right). 
Parameters are $k_B=1$, $\hbar\omega=1$, $\Gamma_1=\Gamma_N=1$, $V=1$, and $g=1$.}
	\label{fig3}
\end{figure}

\section{Dephasing}

The difference between the heat currents of quantum and classical systems can be lifted by adding a dephasing environment to each site of the quantum model. This amounts to introducing an additional term for every site in the master equation~\eqref{eq:ME}:
\be
\mathcal{L}_\text{deph} \rho=\gamma \sum_{k=1}^N \pare{\sigma_k^+\sigma_k^-\rho\sigma_k^+\sigma_k^- - \frac{1}{2}\key{\sigma_k^+\sigma_k^-,\rho}}.
\ee
This term reduces the quantum coherences of the system and thereby progressively transforms the coherent transport into an incoherent, classical one. The transition depends on the parameter $\gamma$. For small values of the dephasing rate the coherent transport is predominant and the transport should be similar to the pure quantum one.
A finite dephasing rate causes an incoherent transport for sufficiently long chains with equal classical probabilities of transition between 
different neighbors, in a similar way to the symmetric simple exclusion process.
For a very high dephasing rate the coherences are reduced dramatically and therefore the transport between neighboring sites is suppressed.

\section{Discussion}
The results for the classical and quantum chain are given in Fig.~\ref{fig2} on a log-log scale. The classical model features a linear dependence in the system size, for high enough values 
of $N$ as expected. The heat current of the quantum case without dephasing is also linear in $N$, but \emph{constant}.
However, when additional dephasing is applied, the heat current is suppressed and now features a size dependence as $1/N$ for sufficiently large values of the dephasing rate~$\gamma$, as confirmed by a numerical analysis of fitting the heat current to a power-law 
 (see Appendix B).
By adding dephasing to the quantum system, we can thus recover the classical $1/N$ dependence of the heat current in the large-$N$ limit.
Phase transitions of this kind have also been observed in low-dimensional models for magnetism; see, e.g., \cite{Znidaric_11}.

In a common interpretation of a dephasing environment, dephasing is caused by fluctuations of the on-site energy of every site. The excited state then effectively forms a band of states that is separated by a gap from the ground state.
Adding dephasing thus effectively recovers the quantum model treated in~\cite{michel:prl05}, and yields the same qualitative result concerning the validity of Fourier's law regarding its dependence on the system size.

Turning from the system size to the temperature-dependence, we find that in the quantum system the heat current features a strong dependence on the temperatures of the heat baths.
Fig.~\ref{fig3} collects the temperature dependencies for both models.
The heat current of the classical system saturates for high values of the temperature, which constitutes a violation of Fourier's law. This is due to the fact that the system has only two levels, which implies a finite heat capacity of the system. Therefore, it cannot transport an arbitrarily large amount of energy, and thus cannot scale linearly with the temperature for a large temperature difference.
The quantum transport features a more intricate behavior. For a high temperature of the hot heat bath, its mean number of excitations and thereby $\gamma_1$ increase, causing a Zeno-type effect that reduces the transport efficiency of the system.
With additional dephasing, the temperature dependence of the heat current of the quantum system approaches a qualitatively similar saturating behavior as in the classical system.

\begin{figure}
	\includegraphics[scale=0.7]{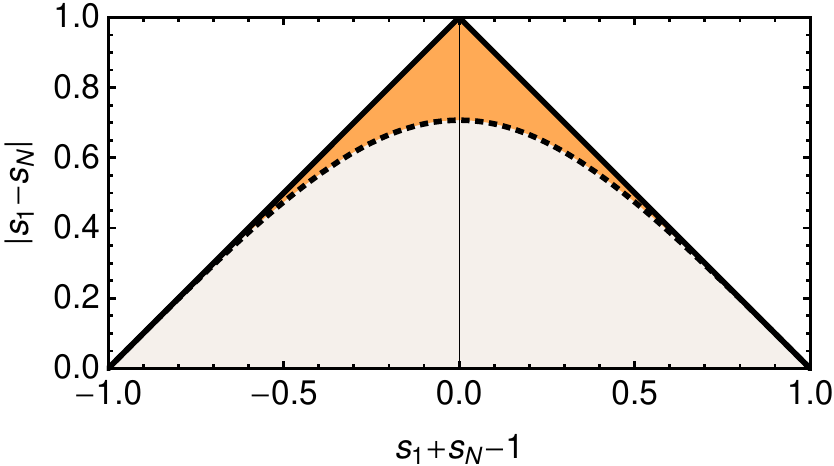}
	\caption{(Color online) Region of parameter space where the steady state may exhibit entanglement for $\gamma_1=\gamma_N$. Within the shaded area of all possible values for $s_1$ and $s_N$, the upper (darker) parameter region indicates values for $s_1$ and $s_N$ where entanglement can occur (dashed boundary not included). That is, only for $s_1$ and $s_N$ in this region do parameters $g$ and $\gamma_1=\gamma_N$ exist such that the steady state is entangled. Entanglement cannot occur for any values of the coupling parameters outside the darker shaded region.}
	\label{fig4}
\end{figure}

\subsection{Disorder}

A relevant point in this respect is the influence of disorder in the system, and the observation that additional noise may unlock the effect of localization in disordered systems for transient transport processes~\cite{aspuru,plenio:njp10}.
However, Figs.~\ref{fig2} and \ref{fig3} show that additional noise due to local dephasing reduces the observed heat current. Although, here, this result is obtained for a chain with uniform on-site energies and inter-site couplings, i.e., in the absence of disorder, we have also numerically investigated disordered chains.
To this extent we have sampled the heat current in chains with $N=5$, with all on-site energies $\hbar\omega_k$ and couplings $g_{kl}$ randomly chosen from a uniform distribution in the interval $[0,1]$.
In 8662 of the 10\,000 disorder samples, we found dephasing to reduce the heat current.
Whenever additional dephasing is found to increase the heat current, the original random configuration exhibited a heat current below average of the entire random ensemble.
We thereby extended what has been observed in the transient case~\cite{scholak:pre11} to the one-dimensional steady-state scenario.

\subsection{Entanglement}

With the perspective of identifying conceivable biological realizations of this transport scenario, an interesting aspect is the question whether entanglement is generated and what role it plays, as addressed in~\cite{Whaley,MT}. Fig.~\ref{fig4} summarizes for which parameters entanglement of the non-equilibrium steady state occurs for a chain with $N=2$ and equal effective bath rates $\gamma_1=\gamma_N$. We find that entanglement can occur, but only in specific regions of the parameter space.
Furthermore, for rates $\Gamma_1=\Gamma_N$, the steady state is never entangled for any choice of bath temperatures and coupling~$g$. A bias in the bath rates, however, may drive the system to an entangled steady state. Depending on the interaction strength between the sites, entanglement may exist for certain range of temperatures.
In contrast to entanglement studies in photosynthesis~\cite{Whaley}, in the present scenario we find that the occurrence of entanglement is not equivalent to, and does not necessarily come with, the mere presence of coherences. It is thus an additional feature.

\section{Conclusions}

To conclude, we have analyzed the energy transfer in a quantum system, formed by a paradigmatic chain of two-level systems, for which we found the heat current in the steady state to be independent of the chain length.
We recover Fourier's law in the quantum-to-classical transition by adding dephasing that destroys quantum coherences.
These results are compared with a purely classical model, the symmetric simple exclusion model, showing that for an 
appropriate value of the dephasing rate the quantum and classical systems exhibit the same qualitative behavior.
It is the coherences in the system that govern the transport properties by design, whereas entanglement may appear independently and in addition for a sufficiently large non-equilibrium.


\bc{\bf ACKNOWLEDGMENTS}\ec
The research was funded by the Austrian Science Fund (FWF) Grants No.\ F04011 and No.\ F04012.
D.M.\ acknowledges funding from the Junta de Andalucia, Projects No.\ FQM-01505 and No.\ FQM-165,
and Spanish MEC-FEDER, Project No.\ FIS2009-08451, together with the Campus de Excelencia Internacional.

\appendix
\section{Purely imaginary coherences}

From the structure of the master equation in Lindblad form~(4), an ordinary linear differential equation, one can directly infer that next-neighbor coherences are imaginary. In Liouville space the equation reads $\dot{\vec{\rho}}=L\vec{\rho}$, where $\vec{\rho}$ is the vector of all matrix elements, which are coupled linearly by the matrix $L$, the Liouvillian.
For our purposes, it is helpful to introduce notation for the matrix elements:
\[
_{12\ldots N}\bra{ik\ldots q}\rho\ket{jl\ldots r}_{12\ldots N} \equiv \rho_{ij,kl,\dotsc,qr},
\]
where indices are grouped by subsystem. The vectors are products of basis vectors of the individual sites with ground state $\ket{0}$ and excited state $\ket{1}$. We thus treat matrix elements with possible indices ``0'' and ``1.''

The Liouvllian $L$ is a sum of three parts, each of which couples certain matrix elements, which yields independent sets of coupled matrix elements. It is possible to distinguish independent sets  by observing general rules of how the Liouvillian couples matrix elements. We formulate these rules by the way indices are transformed by the Liouvillian.

The Lindblad terms $\mathcal{L}_k$ of the Liouvillian inject or extract excitations at the terminal sites of the chain. Thereby, they transform matrix elements into one another that differ only by a pair of ``00'' and ``11'' indices of the first or last subsystem, e.g., $\rho_{\boldsymbol{00},01} \leftrightarrow \rho_{\boldsymbol{11},01}$. This constitutes a change of the total number of indices ``0'' and ``1'' by two, hence leaving the respective total number of indices ``0'' and ``1'' even or odd.
Since $H$ commutes with the excitation number operator, the commutator that appears in the Liouvillian leaves the total number of excitations invariant and hence couples only matrix elements with the same number of indices ``0'' and ``1,'' respectively. The coherent dynamics captures the exchange of excitations between neighboring sites and, in terms of matrix elements, couples those that can be transformed into each other by exchanging a ``0'' and a ``1'' index between neighbors, while maintaing the relative index position, i.e., left and right indices are transformed within themselves, e.g., $\rho_{\boldsymbol{0}1,\boldsymbol{1}0} \leftrightarrow \rho_{\boldsymbol{1}1,\boldsymbol{0}0}$.
This implies that the ground state $\rho_{00,00,\dotsc}$ is coupled to all other populations, i.e., matrix elements with indices of the form $\rho_{ii,jj,\dotsc}$, and only to those coherences that contain an equal number of indices ``1'' on the left and right.
The remaining matrix elements form an independent closed set of equations, whose steady-state solution is therefore the trivial solution, where all matrix elements vanish.
(The set of equations that includes the populations is not solved by the trivial solution in the steady state because it is subject to the boundary condition $\Tr\rho=1$.) Note that the diagonal of $L$ contains only coefficients with negative real parts, meaning that all matrix elements would decay to zero if not sufficiently maintained by a positive contribution due to another element. A population is coupled to a next-neighbor coherence, e.g., $\rho_{11,00} \leftrightarrow \rho_{01,10}$, with a coupling $\pm i g$ such that (real) populations pump the imaginary part of the next-neighbor coherences, and vice versa. The contributions to and from the real part of the latter cancel, leading to their decay. In longer chains ($N>2$) next-neighbor coherences are also coupled to next-to-nearest-neighbor coherences, e.g., $\rho_{01,10,00} \leftrightarrow \rho_{01,00,10}$, with the same factor $\pm i g$ thus coupling the imaginary (real) part of the former to the real (imaginary) part of the latter.
Therefore, the imaginary and real parts the of next-neighbor coherences belong again to different and independent sets of coupled differential equations. The real parts have the trivial solution, whereas the imaginary part is non-zero in the steady state.

\section{Fit of dephasing numerics}
To analyze the behavior of the quantum system under the effect of dephasing, we take the logarithm of~(2) and obtain a linear dependence between $\log J$ and $\log N$: 
\[
\ln J= \ln \pare{ -c\Delta T} + (\alpha-1) \ln N.
\]
By a linear regression of the numerical data of Fig.~2, for $\gamma=5$ we obtain a value $\alpha=0.0242$ with a regression coefficient $R=0.999\,985\,1$. The small discrepancy with Fourier's law ($\alpha=0$) is due to the finite size of the system.

\end{document}